\documentclass[twocolumn,prl,floats,preprintnumbers,showpacs]{revtex4}

\usepackage[dvips]{graphicx}
\usepackage{amsmath,amssymb}

\begin{document}

\def\lsim{\; \raise0.3ex\hbox{$<$\kern-0.75em
      \raise-1.1ex\hbox{$\sim$}}\; }
\def\gsim{\; \raise0.3ex\hbox{$>$\kern-0.75em
      \raise-1.1ex\hbox{$\sim$}}\; }

%
\newcommand{\nc}{\newcommand}
\newcommand{\comment}[1]{}

\nc{\half}{\textstyle\frac{1}{2}}
\nc{\fsky}{f_{\rm sky}}
\nc{\fwhm}{\theta_{\rm fwhm}}
\nc{\fwhmc}{\theta_{{\rm fwhm},c}}
\nc{\nadi}{n_{\rm ad1}}
\nc{\nadis}{n_{\rm ad2}}
\nc{\niso}{n_{\rm iso}}
\nc{\ncor}{n_{\rm cor}}
\nc{\fiso}{f_{\rm iso}}
\nc{\wmap}{\textsc{wmap} }
\nc{\camb}{\textsc{camb} }
\nc{\R}{{\cal{R}}}

\nc{\GeV}{\mbox{ GeV}}
\nc{\MeV}{\mbox{ MeV}}
\nc{\keV}{\mbox{ keV}}
\nc{\etal}{{\it et al.}}

\title{Correlated adiabatic and isocurvature CMB fluctuations in the wake of
  the WMAP}
\author{Jussi V\"{a}liviita}
\email{jussi.valiviita@helsinki.fi}

\author{Vesa Muhonen}
\email{vesa.muhonen@helsinki.fi}

\affiliation{Department of Physical Sciences, University of Helsinki, and
  Helsinki Institute of Physics, P.O. Box 64, FIN-00014 University of
  Helsinki, Finland}

\date{September 30, 2003}

\begin{abstract}
  In general correlated models, in addition to the usual adiabatic
  component with a spectral index $\nadi$ there is another adiabatic component
  with a spectral index $\nadis$ generated by entropy perturbation during
  inflation. We extend the analysis of a correlated mixture of adiabatic and
  isocurvature CMB fluctuations of the \wmap group, who set the two adiabatic
  spectral indices equal. Allowing $\nadi$ and $\nadis$ to vary independently
  we find that the \wmap data favor models where the two adiabatic components
  have opposite spectral tilts. Using the \wmap data only, the $2\sigma$ upper
  bound for the isocurvature fraction $\fiso$ of the initial power spectrum
  at $k_0=0.05$ Mpc$^{-1}$ increases somewhat, e.g., from 0.76 of
  $\nadis = \nadi$ models to 0.84 with a prior $\niso < 1.84$ for the
  isocurvature spectral index. We also comment on a possible degeneration
  between the correlation component and the optical depth
  $\tau$. Moreover, the measured low quadrupole in the TT angular power
  could be achieved by a strong negative correlation, but then one needs
  a large $\tau$ to fit the TE spectrum.

\end{abstract}
\pacs{98.70.Vc, 98.80.Cq}
\preprint{HIP-2003-21/TH}
\preprint{Phys.\ Rev.\ Lett.\  {\bf 91}, 131302 (2003)}
\maketitle
%
%
\textbf{Introduction.}
The first studies of mixed initial conditions for
density perturbations in the light of measured cosmic microwave background
(CMB) angular power assumed the adiabatic and isocurvature components to be
\emph{uncorrelated} \cite{Enqvist:2000hp,Enqvist:1999vt,Pierpaoli:1999zj}.
About the same time it was pointed out that inflation with more than one
scalar field may lead to a \emph{correlation} between the adiabatic and 
isocurvature perturbations \cite{Langlois:dw}. If the trajectory in the field 
space is curved during inflation, the entropy perturbation generates an adiabatic
perturbation that is fully correlated with the entropy perturbation
\cite{Garcia-Bellido:1995fz,Gordon:2000hv,Amendola:2001ni,Gordon:2001ph}.  In
addition, there is also the usual adiabatic perturbation created, e.g., by
inflaton fluctuations.  Thus, in the final angular power spectrum, one could
have four different components: (1) the usual independent adiabatic component, (2)
a second adiabatic component generated by the entropy perturbation during
inflation, (3) an isocurvature component, and (4) a correlation between the second
adiabatic and the isocurvature component. In this Letter we assume power laws
for the initial power spectra of these components and denote their spectral
indices by $\nadi$, $\nadis$, $\niso$, and $\ncor$, respectively.  Only three
of these are free parameters, since, e.g., $\ncor = (\nadis + \niso)/2$.

Although \emph{pure} isocurvature models have been ruled out
\cite{Enqvist:2001fu} after the clear detection of the second acoustic peak
\cite{Netterfield:2001yq}, the \emph{correlated mixture} of adiabatic and
isocurvature fluctuations still remains as an interesting possibility.  In
\cite{Amendola:2001ni,Trotta:2001yw} angular power spectra have been
calculated for correlated models and compared to the CMB data, but the spectral
indices have either been fixed or set equal, $\nadi = \nadis = \niso = \ncor$.
This is not necessarily well motivated theoretically. For example, if the
entropy field is slightly massive during inflation, then $\nadi \lsim 1.0 <
\niso$ in most models.

Recently, the Wilkinson Microwave Anisotropy Probe (\textsc{wmap}) accurately
measured the temperature anisotropy spectrum up to the second acoustic peak
\cite{WMAP1} and also the TE cross-correlation \cite{WMAP8}, which plays an
important role in constraining cosmological models. The \wmap group considered the
possibility of mixed models in \cite{WMAP11} where, in order to simplify the
analysis, they set the two adiabatic spectral indices equal, $\nadis = \nadi$.
They found that a correlated mixture of adiabatic and isocurvature
fluctuations does not improve the fit to the data.

However, we would rather expect the second adiabatic spectral index to be
close to the isocurvature one, $\nadis \approx \niso$, since both of these
fluctuation components have been generated by the entropy perturbation during
inflation. In this Letter we study a correlated mixture of the adiabatic and cold
dark matter isocurvature fluctuations relaxing the ``\wmap condition'' by
letting $\nadis \neq \nadi$. We show that the data clearly allows this and, e.g.,
the upper bound for the isocurvature fraction, $\fiso$, slightly weakens. In this
preliminary analysis we use the \wmap data set only, but allow $\nadi$,
$\nadis$, and $\niso$ and the amplitudes of different components to vary
independently to see if there are any interesting effects.  A more thorough
analysis including other CMB and large scale structure data will be
presented in \cite{MuhonenValiviita}.

\textbf{Dealing with correlation.}  In the following we consider linear
perturbation theory in the spatially flat ($\Omega = 1$) universe. For
simplicity, we assume inflation with two scalar fields only. Following
\cite{Gordon:2000hv} we denote the inflaton field by $\phi$ and the other
scalar field by $\chi$.

The transformation of the comoving curvature perturbation $\hat{\cal R}$ and
the entropy perturbation $\hat{\cal S}$ from the time of the Hubble length exit
during inflation to the beginning of radiation dominated era is of the form
\cite{Amendola:2001ni}
\begin{equation}
  \begin{pmatrix}
    \hat{\cal R}_{{\rm rad}}(k) \\ 
    \hat{\cal S}_{{\rm rad}}(k)
  \end{pmatrix}
  =
  \begin{pmatrix}
    1 & T_{{\cal R}{\cal S}}(k) \\ 
    0 & T_{{\cal S}{\cal S}}(k)
  \end{pmatrix}
  \begin{pmatrix}
    \hat{\cal R}_*(k) \\ 
    \hat{\cal S}_*(k)
  \end{pmatrix}\,,
  \label{GeneralTransfer}
\end{equation}
where the transfer functions $T_{{\cal R}{\cal S}}(k)$ and $T_{{\cal S}{\cal
    S}}(k)$ carry all the information about the evolution of the perturbations.
They are obtained by solving \emph{numerically} the equations of motion for
the adiabatic and entropy perturbations during inflation and reheating.
Almost all the way from the generation of classical perturbations during
inflation to the beginning of radiation dominated era
the cosmologically interesting perturbation modes are super-Hubble, $k \ll aH$.
Then the evolution of
perturbations is practically $k$ independent \cite{Gordon:2000hv}.  However,
reheating process may change this conclusion.  Here we assume that
$T_{\mathcal{S}\mathcal{S}}$ is only weakly $k$ dependent. The easiest way to
determine $ T_{\mathcal{R}\mathcal{S}}$ is to use the equation $\dot {\cal R} =
2H\dot\theta\delta s/\dot\sigma$, where $\sigma$ is the background homogeneous
adiabatic field, $\delta s$ is the entropy field perturbation [$\hat{\delta s}
= -(2V_s/3\dot\sigma^2)\hat {\cal S}$], and $\theta$ is the angle between the
inflaton field $\phi$ and the trajectory $(\phi(t),\chi(t))$ in the field
space as defined in \cite{Gordon:2000hv}.  Writing $\hat{\mathcal{R}}_{2\,\rm rad}$
for the entropy generated curvature perturbation $T_{{\cal R}{\cal S}}(k)
\hat{\cal S}_*(k)$ we get
\begin{equation}
  \hat{\cal R}_{2\,\mathrm{rad}}(k)
  =  \int_{t_*(k)}^{t_{\mathrm{rad}}} dt
  \frac{2H(t)}{\dot\sigma(t)}
  \dot\theta(t)\hat{\delta s}(k,t)\,,
  \label{EntropyGenAdi}
\end{equation}
where $t_*(k)$ is the time when mode $k$ becomes super-Hubble. If the fields
and the angle $\theta$ are evolving slowly during the generation of perturbations,
then the main $k$ dependence of $\hat{\cal R}_{2\,\rm rad}$ comes from the
initial $k$ dependence of $\hat{\delta s}(k,t_*)$, which is very close to that
of $\hat{\cal S}_{{\rm rad}}$, assuming $T_{{\cal S}{\cal S}}$ to be also
weakly $k$ dependent. In this situation the $k$ dependence of the integration
lower bound $t_*(k)$ translates into a small correction to the $k$ dependence
of $\hat{\cal R}_{2\,\rm rad}$ when compared to $\hat{\cal S}$. This suggests
that $\hat{\cal R}_{2\,\rm rad}$ and $\hat{\cal S}$ have nearly identical $k$
dependence. On the other hand,
if the entropy field $\delta s$ is massive, it leads to $\hat{\cal
  S}_{\rm rad}$ that increases as a function of $k$ while, at the same time due to
slow roll, the usual adiabatic component, $\hat{\cal R}_{1\,\rm rad} =
\hat{\cal R}_*$ generated by inflaton fluctuations, can be nearly scale
independent as in the simplest inflationary scenarios.

We define the correlation ${\cal C}_{xy}(k)$ between two perturbation
quantities $x$ and $y$, which in our case are $\cal R$ for the adiabatic and $\cal
S$ for the isocurvature fluctuation, at the beginning of radiation dominated
era by the formula
\begin{equation}
  \big\langle x(\vec k) y^*(\vec k') \big\rangle \big\vert_{\rm rad}
  = \tfrac{2\pi^2}{k^3}\, {\cal C}_{xy}(k) \, \delta^{(3)}(\vec k- \vec k') \,.
  \label{correlationC}
\end{equation}
The angular power spectrum induced by the ${\cal C}_{xy}$ will be
\begin{equation}
  C_{xy\,l} = \int \frac{dk}{k}{\cal C}_{xy}(k)
  g_{x\,l}^{(T/E/B)}(k)  g_{y\,l}^{(T/E/B)}(k)\, ,
  \label{AngularPower}
\end{equation}
where $g_l$ is the transfer function that describes how an initial perturbation
evolves to a presently observable temperature ($T$) or polarization ($E$- or
$B$-mode) signal at the multipole $l$.

Assuming that everything changes slowly in time and is weakly $k$ dependent
in a sense described after the equation (\ref{EntropyGenAdi}), the  end
result of (\ref{GeneralTransfer}) is well approximated by the power laws
\begin{equation}
  \begin{split}
    \bigl( \tfrac{k^3}{2\pi^2} \bigr)^{\frac{1}{2}}
    \hat{\cal R}_{\rm rad} &= A_r \bigl(\textstyle{\frac{k}{k_0}}
    \bigr)^{n_1} 
    \hat{a}_r(\vec k) + A_s
    \bigl(\textstyle{\frac{k}{k_0}}\bigr)^{n_3} \hat{a}_s(\vec k)\,,
    \\
    \bigl( \tfrac{k^3}{2\pi^2} \bigr)^{\frac{1}{2}}
    \hat{\cal S}_{\rm rad} &= B  \bigl(\textstyle{\tfrac{k}{k_0}}
    \bigr)^{n_2}
    \hat{a}_s(\vec k) \,,
  \end{split}
  \label{InitialFluct}
\end{equation}
where $\hat a_r$ and $\hat a_s$ are Gaussian random variables obeying
\begin{equation*}
  \langle \hat a_r \rangle = 0, \ \langle \hat a_s \rangle = 0, \
  \langle \hat a_r(\vec k) \hat a_s^*(\vec k')\rangle 
  = \delta_{rs}\delta^{(3)}(\vec k - \vec k')\,.
\end{equation*}
$A_r$, $A_s$, and $B$ are the amplitudes of the usual adiabatic, the entropy
generated second adiabatic, and the isocurvature component, respectively.
We define
$\tilde{k} = k/k_0$, where $k_{0} = 0.05$ Mpc$^{-1}$ is the wave number of a
reference scale. As explained, we expect $n_2 \simeq n_3$ in the case where
$H\dot\theta/\dot\sigma$ is nearly constant and much less than one during the
generation of perturbations. If this condition is not satisfied, then $n_2$ and
$n_3$ can take completely different values. General expressions for the
spectral indices in terms of the slow roll parameters are derived in
\cite{Bartolo:2001rt}.  Actually, even the power law spectra (\ref{InitialFluct})
may be bad
approximations. E.g., in double inflation numerical studies show that
the perturbations can be strongly scale dependent \cite{Tsujikawa:2002qx}.

Inserting (\ref{InitialFluct}) into (\ref{correlationC}), the
autocorrelations become
\begin{equation}
  {\cal C}_{\cal RR} = A_r^2 \tilde k^{2n_1} + A_s^2 \tilde k^{2n_3}
  \;\; \text{and} \;\;
  {\cal C}_{\cal SS} = B^2 \tilde{k}^{2n_2}\,,
  \label{CRRCSS}
\end{equation}
while the cross-correlation between the adiabatic and isocurvature fluctuations is
\begin{equation}
  {\cal C}_{\cal RS}(k) 
  = {\cal C}_{\cal SR}(k)
  = A_s B \tilde{k}^{n_3+n_2}.
  \label{CRS}
\end{equation}
Substituting (\ref{CRRCSS}) and (\ref{CRS}) into
(\ref{AngularPower}) and noting that
the present total angular power is
\begin{equation*}
  C_l = C_{{\cal R R}\,l} + C_{{\cal S S}\,l} + C_{{\cal R S}\,l} 
  + C_{{\cal S R}\,l}\,,
\end{equation*}
we get for the temperature angular power spectrum
\begin{equation}
  \begin{split}
    C_l^{TT} & =  \int\!\frac{dk}{k}\!\Bigl[
      A_r^2 (g_{{\cal R}\,l}^T )^2 \tilde k^{2n_1}
      + A_s^2 (g_{{\cal R}\,l}^T )^2 \tilde k^{2n_3}
    \\
    & \quad+ B^2 (g_{{\cal S}\,l}^T )^2 \tilde k^{2n_2}
      + 2 A_s B g_{{\cal R}\,l}^T  g_{{\cal S}\,l}^T  \tilde k^{n_3+n_2}
    \Bigr],
  \end{split}
  \label{CTT}
\end{equation}
and for the TE cross-correlation spectrum
\begin{multline}
  C_l^{TE} =  \int\!\frac{dk}{k}\!\Bigl[
    A_r^2 g_{{\cal R}\,l}^T  g_{{\cal R}\,l}^E  \tilde k^{2n_1}
    + A_s^2 g_{{\cal R}\,l}^T  g_{{\cal R}\,l}^E  \tilde k^{2n_3}
  \\
  + \!
    B^2 g_{{\cal S}\,l}^T  g_{{\cal S}\,l}^E   \tilde k^{2n_2}
    \! + \!
    A_s B \left( 
      g_{{\cal R}\,l}^T  g_{{\cal S}\,l}^E  
      \! + \! 
      g_{{\cal S}\,l}^T  g_{{\cal R}\,l}^E  \right) \tilde k^{n_3+n_2}
  \Bigr].
  \label{CTE}
\end{multline}
Above we defined $n_{1}$, $n_{2}$, and $n_{3}$ so that for the scale free case
they are zeros. To match the historical convention, we define new spectral
indices as follows: $\nadi - 1 = 2n_1$, $\niso - 1 = 2n_2$, and $\nadis - 1 =
2n_3$.

The amplitudes are not yet in a convenient
form in (\ref{CRRCSS}) and (\ref{CRS}). The overall adiabatic amplitude at the
reference scale $k_0$ is $A^2 = A_r^2 + A_s^2$. Using this, the adiabatic
initial power spectrum can be written as
\begin{equation}
  {\cal C_{RR}} = A^2\left[ (1-Y^2) \tilde k^{\nadi-1}
    + Y^2 \tilde k^{\nadis-1} \right]\,,
  \label{CRR}
\end{equation}
where $Y^2 = A_s^2/A^2$ so that $0 \le Y^2 \le 1$.
Following \cite{WMAP11} we define the isocurvature
fraction by $\fiso^2 = (B/A)^2$ and obtain
\begin{equation*}
  {\cal C_{SS}} = A^2 \fiso^2 \tilde k^{\niso-1}\,.
\end{equation*}
The correlation amplitude is $A_s B = A^2 (B/A) (A_s/A) = A^2 \fiso {\rm
  sign}(B)\sqrt{Y^2}$ from \eqref{CRS}.  Without loss of generality, the total
angular power spectrum can now be written as
\begin{equation*}
  C_l \! = \! A^2\!\big(\sin^2\!\!\Delta C_l^{\rm ad1}
  + \cos^2\!\!\Delta C_l^{\rm ad2}
  + \fiso^2 C_l^{\rm iso} + \fiso\!\cos\!\Delta C_l^{\rm cor}\big),
\end{equation*}
where $A^2 > 0$ is the overall amplitude, $0 \le \Delta \le \pi$, and $\fiso \ge
0$.  $C_l^{\rm ad1}$, $C_l^{\rm ad2}$, $C_l^{\rm iso}$, and $C_l^{\rm cor}$ are
calculated by our modified version of \camb \cite{camb} for each cosmological
model from (\ref{CTT}) or (\ref{CTE}) keeping $A_r = A_s = B = 1$. For
example, $C_l^{\rm ad TT}$ is given by the first term in the integral
(\ref{CTT}) and the last term of integral (\ref{CTE}) gives $C_l^{\rm cor
  TE}$.

If $\nadi$ and $\nadis$ are nearly equal or amplitude $Y^2$ is close to zero
or one, the adiabatic power spectrum in (\ref{CRR}) can well be approximated
by a single power law ${\cal C_{RR}} = D \tilde k^{n_{\rm ad}-1}$, where $D$ is
the amplitude. However, in the general case, an attempt to write the term in
square brackets in (\ref{CRR}) in terms of a single power law leads to a
strongly scale dependent spectral index $n_{\rm ad}(\tilde k) - 1 = d\ln{\cal
  C_{RR}}(\tilde k)/d\ln \tilde k$.  The first derivative of this is always
non-negative:
\begin{equation*}
  \frac{dn_{\rm ad}(\tilde k)}{d\ln \tilde k}
  = \frac{(1-Y^2)Y^2(\nadi - \nadis)^2 \tilde k^{\nadi+\nadis}}{[(1-Y^2)
    \tilde k^{\nadi} + Y^2 \tilde k^{\nadis}]^2}.
\end{equation*}
The \wmap group observed that the combined CMB and other cosmological data
favor a running spectral index with a \emph{negative} first derivative. Thus
one would expect that the data disfavor models where $\nadi \neq \nadis$,
since this evidently leads to a positive first derivative of $n_{\rm ad}$.
However, the correlation power spectra ${\cal C_{RS}}$ and ${\cal C_{SR}}$
may well balance the situation so that a more comprehensive analysis
\cite{MuhonenValiviita} is needed.

\textbf{Technical details of analysis.}  In this preliminary analysis we
consider spatially flat ($\Omega = 1$) universe and use a
coarse grid method leaving a sophisticated Monte Carlo analysis
\cite{montecarlo} for future work \cite{MuhonenValiviita}. We concentrate on
the neighborhood of the best-fit adiabatic model found in \cite{WMAP9}.
Naturally, this favors pure adiabatic models, but our primary interest is not
to do full confidence level cartography here. Instead we study whether
relaxing the \wmap constraint $\nadis = \nadi$ has any interesting effects. Hence,
we scan the following region of the parameter space: reionization optical depth
$\tau$ = 0.11 -- 0.19 (step 0.02), vacuum energy density parameter
$\Omega_\Lambda$ = 0.69 -- 0.77 (0.02), baryon density $\omega_b$ = 0.021 --
0.025 (0.001), cold dark matter density $\omega_c$ = 0.10 -- 0.18 (0.02),
$\nadi$ = 0.73 -- 1.27 (0.03), $\nadis$ = 0.55 -- 1.84 (0.03), $\niso$ = 0.55
-- 1.84 (0.03), $\fiso$ = 0.0 -- 1.2 (0.04), $\cos\Delta$ = -1.0 -- 1.0
(0.04).  The best overall amplitude $A^2$ is found by maximizing the
likelihood for each model. We stress that, as in any similar analysis, the
choice of the grid is a top-hat prior in itself.

Since the likelihood code offered by the \wmap group \cite{WMAP7,WMAP8,WMAP10}
is far too slow for a grid method, we are able to use only the diagonal elements
of the Fisher matrix when calculating the likelihoods $\mathcal{L}$.  Ignoring
the off-diagonal terms increases the effective $\chi^2 \equiv
-2\ln\mathcal{L}$ by about 4 from 1428 for well-fitted models, but since
this effect is common to all models, it has only a small effect on the
confidence level plots.  However, we point out that the results presented here
are mostly qualitative in nature.

\textbf{Results.}  From the likelihoods on the $(\nadis,\nadi)$ plane in Fig.~1(a),
marginalized by integrating over all the other parameters, we see that the
data do not especially favor $\nadis = \nadi$. Clearly most of the $2\sigma$
allowed models are in the regions where one of the adiabatic spectral indices
is larger than 1.0, the other being less than 1.0.  Hence the \wmap data
favor models where the adiabatic components have the opposite spectral tilts.
Using the full Fisher matrix of \wmap our best-fit model gives $\chi^2 =
1427.8$ while the best-fit $\nadis = \nadi$ model has $\chi^2 = 1428.0$.  For
comparison, our best-fit pure adiabatic model has $\chi^2 = 1429.0$.  So,
allowing for a correlated mixture improves the fit slightly. However, in pure
adiabatic models the number of degrees of freedom is $\nu = 1342$ while in
correlated mixed models we have four additional parameters leading to $\nu =
1338$.
Thus the goodness-of-fit of pure adiabatic models is about the same as that of
mixed models:
$\chi^2/\nu = 1.065$ for adiabatic and $\chi^2/\nu = 1.067$ for mixed models.

Figure 1(b) shows that the isocurvature
spectral index $\niso$ is not limited from above.
To get a constraint one would need to include some large scale structure
data, which we expect to give about $\niso \lsim 1.8$ \cite{WMAP11}
motivating our prior $\niso < 1.84$.
When the isocurvature fraction $\fiso$ (at $k_0 = 0.05$ Mpc$^{-1}$,
corresponding the multipole $l_{\mathrm{eff}}\approx 700$) is large, the data favor
large $\niso$, i.e., positively tilted isocurvature spectrum in order to get
less power at the smallest multipoles $l$. We show also by dashed lines how
the \wmap restriction $\nadis=\nadi$ modifies the contours. The difference is
clear in $1\sigma$ region but $2\sigma$ regions are nearly identical. From
one-dimensional, slightly non-Gaussian, marginalized likelihood function of
$\fiso$ we find a $2\sigma$ upper bound for the isocurvature fraction, $\fiso
\lsim 0.84$. With the restriction $\nadis=\nadi$, the bound would be about
$\fiso \lsim 0.76$.

\begin{figure*}[t]
  \centering
  \includegraphics[width=0.36\textwidth]{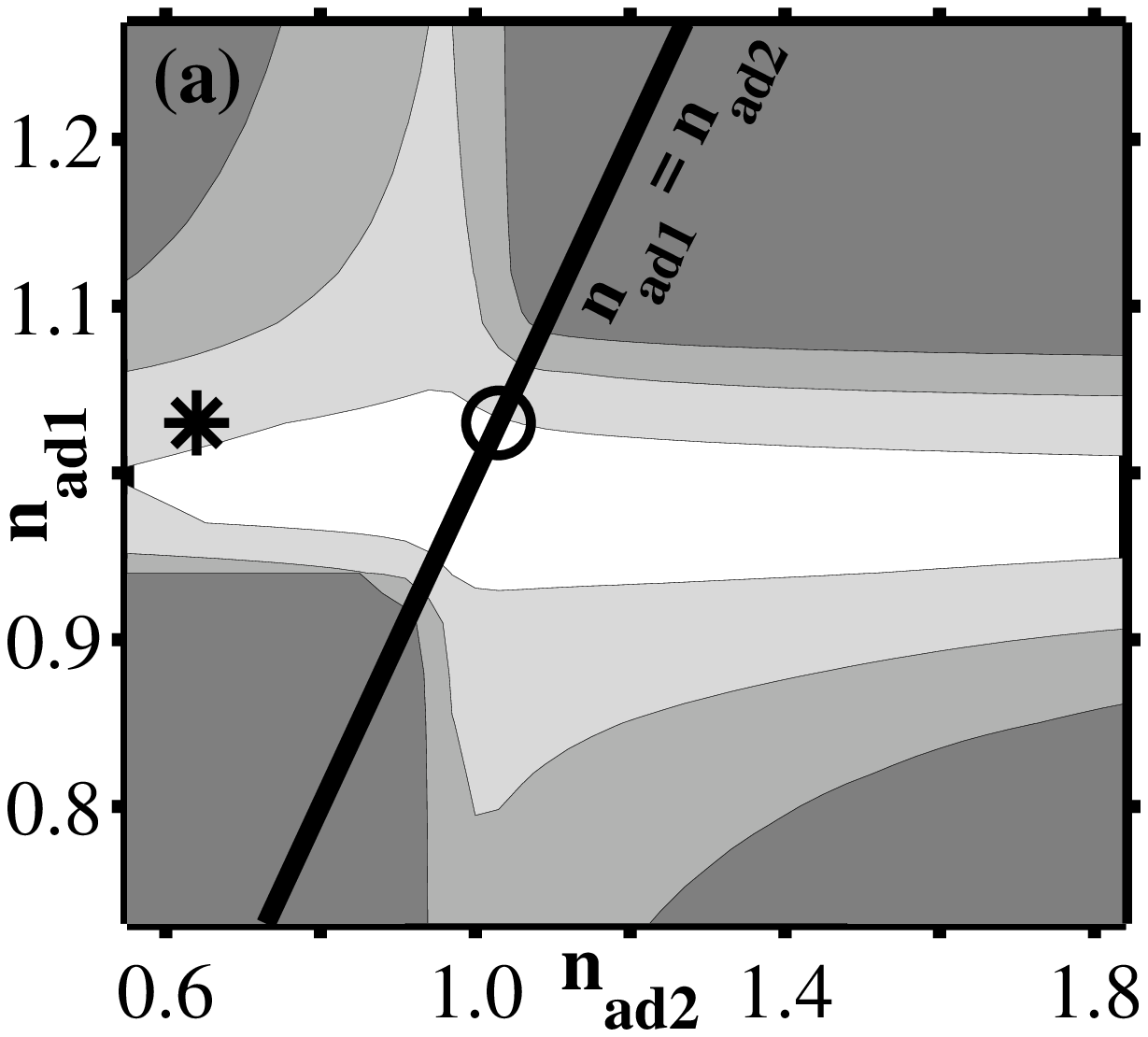}\hspace{0.8cm}
  \includegraphics[width=0.36\textwidth]{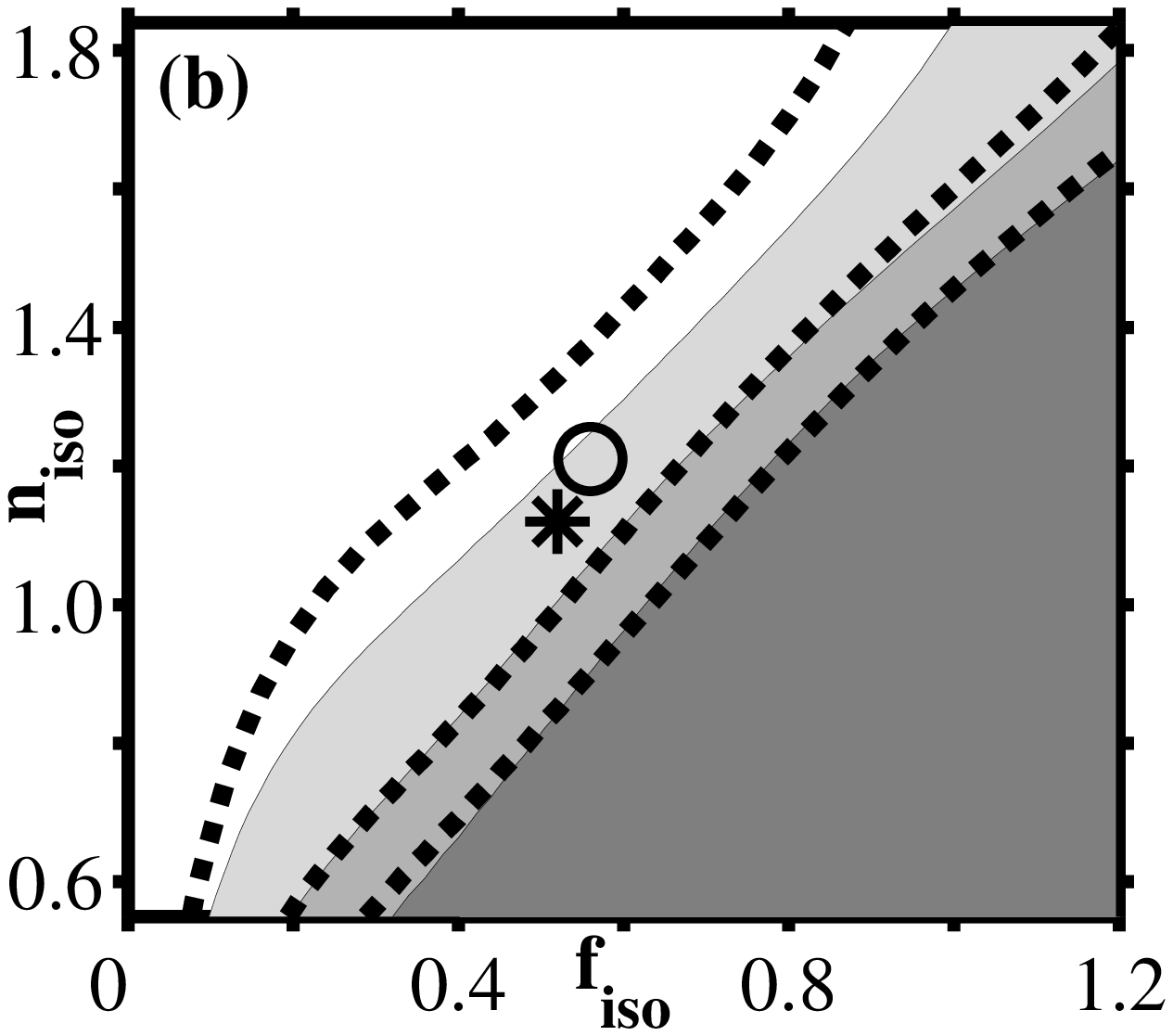}\vspace{0.47cm}
  \includegraphics[width=0.36\textwidth]{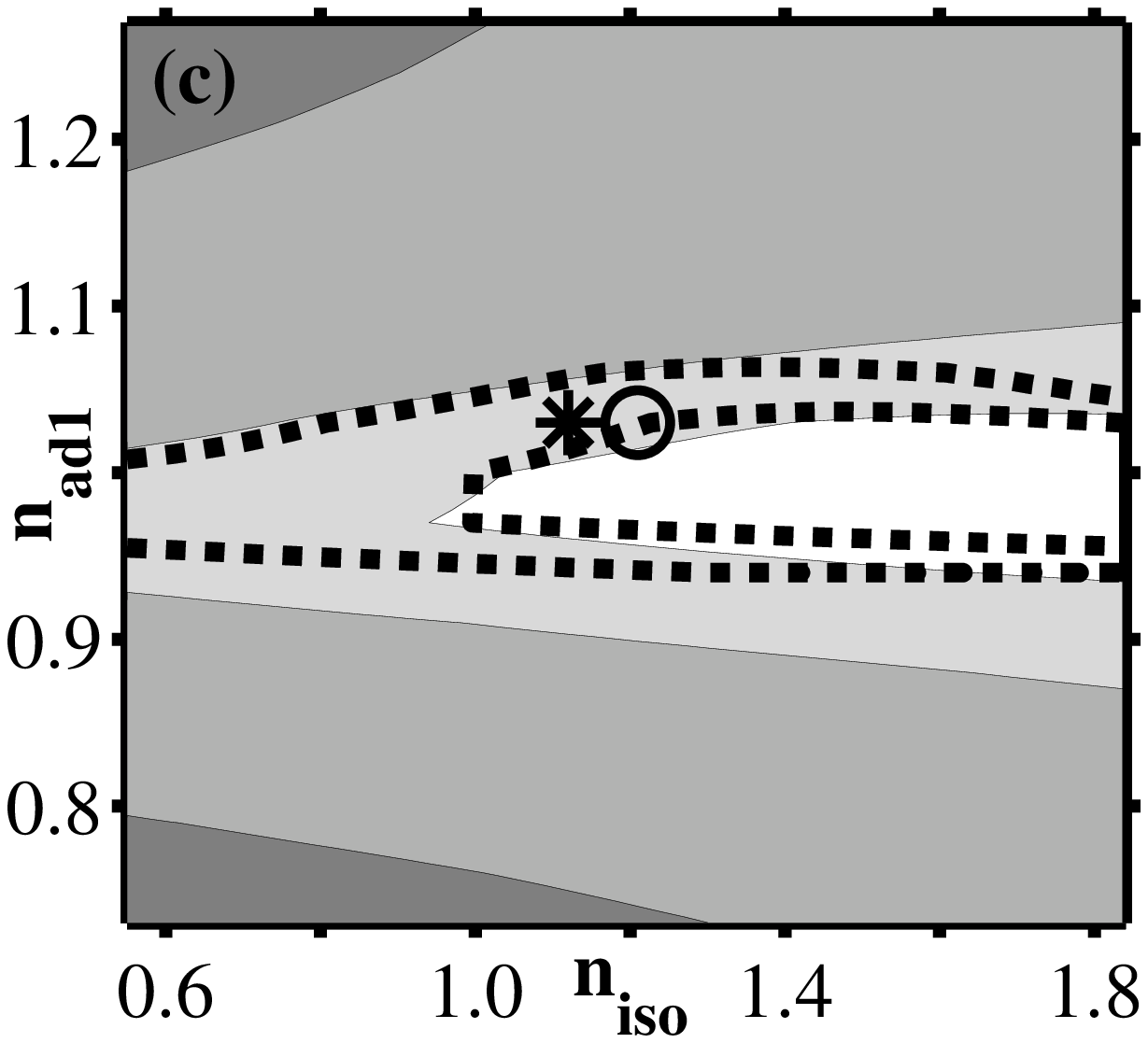}\hspace{0.8cm}
  \includegraphics[width=0.36\textwidth]{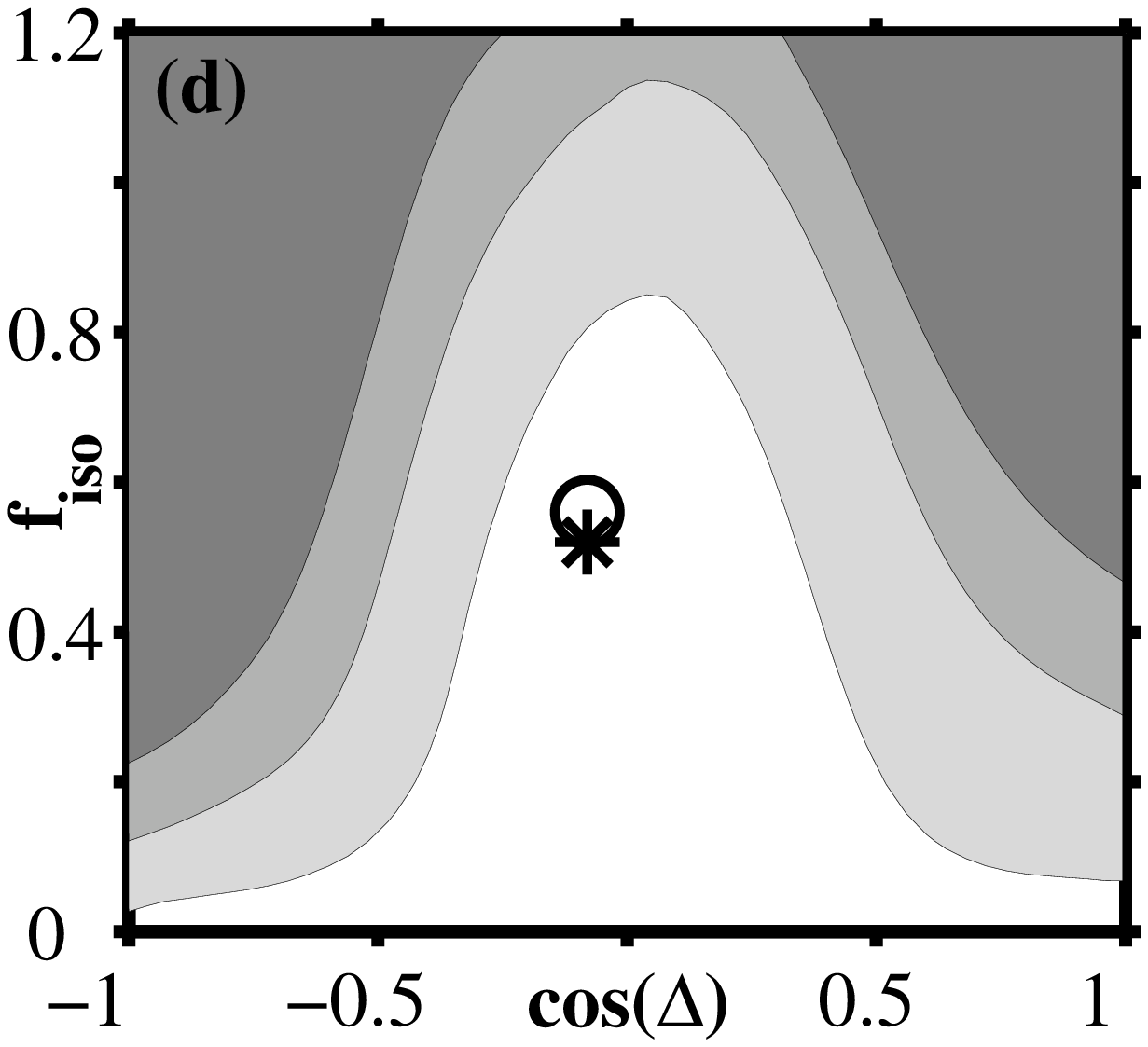}\vspace{0.47cm}
  \includegraphics[width=0.36\textwidth]{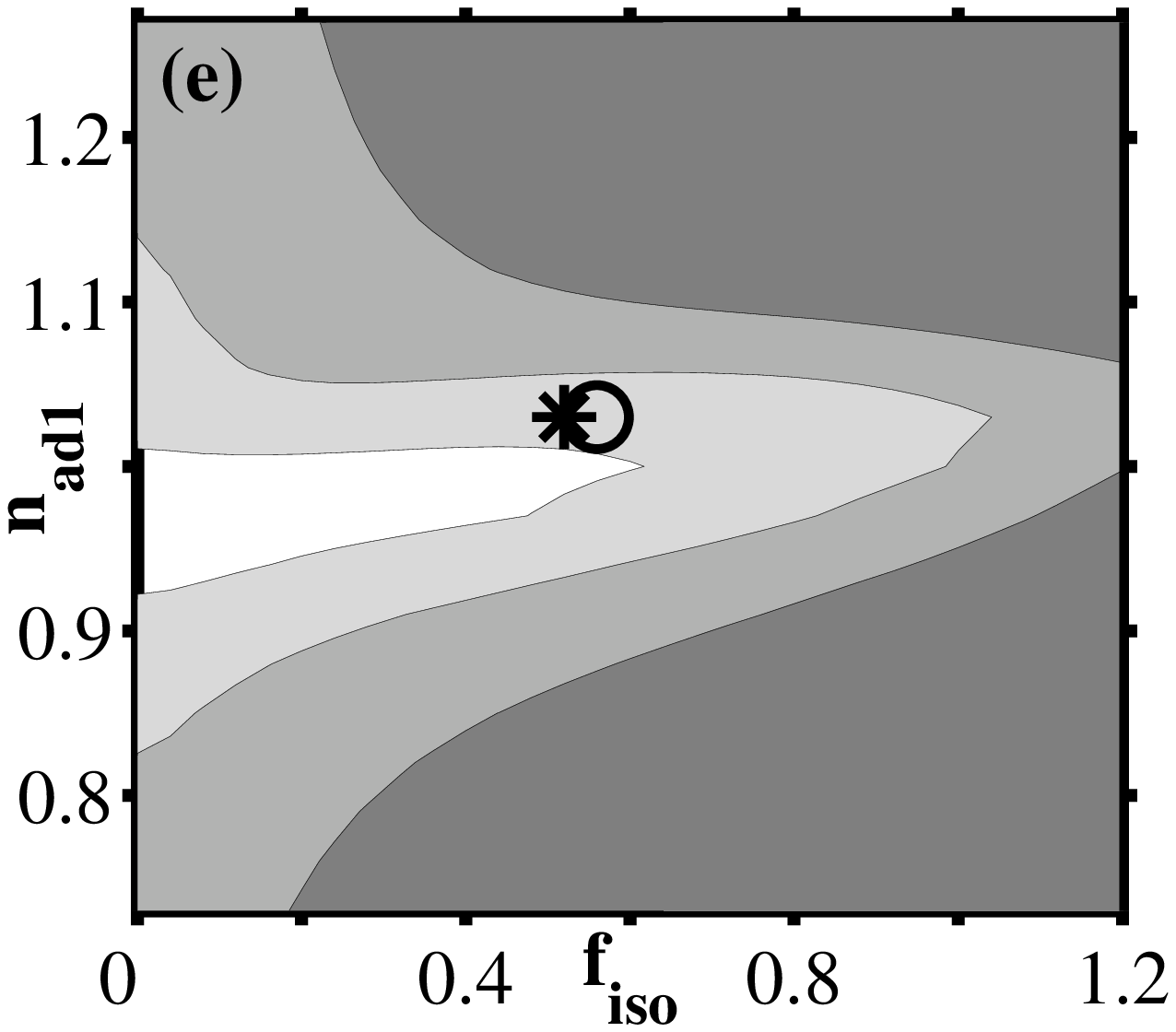}
\caption{The 68.3\%/$1\sigma$ (white),
    95.4\%/$2\sigma$ (light gray),
    99.7\%/$3\sigma$ (medium gray),
    and more than $3\sigma$ (dark gray)
    confidence levels for our general models.
    The best-fit model 
($\tau$, $\Omega_\Lambda$, $\omega_b$, $\omega_c$, $\nadi$, $\nadis$, 
$\niso$, $\fiso$, $\cos\Delta$)
= (0.13, 0.73, 0.025, 0.12, 1.03, 0.64, 1.12, 0.52, -0.08)
is marked by an asterisk
    ($\ast$) and the best-fit $\nadis=\nadi$ model by a circle ($\circ$).
    The dashed lines in (b) are confidence levels for
    $\nadis=\nadi$ models
    and in (c) they indicate
    $1\sigma$ and $2\sigma$ regions for uncorrelated models, i.e.,
    $\cos\Delta = 0$.} 
\end{figure*}

Figure 1(c) should be compared with the results obtained for an uncorrelated
mixture of adiabatic and isocurvature fluctuations in \cite{Enqvist:2000hp}.
Although qualitatively similar, the $1\sigma$ and $2\sigma$ regions of
uncorrelated models are much smaller than in \cite{Enqvist:2000hp} due
to improved accuracy of the data. We also notice that allowing for
correlated models significantly enlarges the $2\sigma$ allowed region
in the parameter space.

Since the correlation amplitude is $\fiso\cos\Delta$,
it is natural that for a high isocurvature fraction $\fiso$ the data
prefer smaller $|\cos\Delta|$, which
is evident in Fig.~1(d). Comparing one-dimensional
marginalized likelihoods for $\nadi$ in the pure adiabatic case and
in the correlated models we find that allowing for a correlation
does not affect much the usual adiabatic spectrum, which is
nearly scale free: $\nadi = 0.97 \pm 0.06$ (pure adiabatic),
$\nadi = 0.98 \pm 0.07$ (correlated models). However, Fig.~1(e) shows
that increasing the isocurvature fraction leads to slightly stricter
bounds for $\nadi$.    

\begin{figure*}[t]
  \centering
  \includegraphics[width=0.38\textwidth]{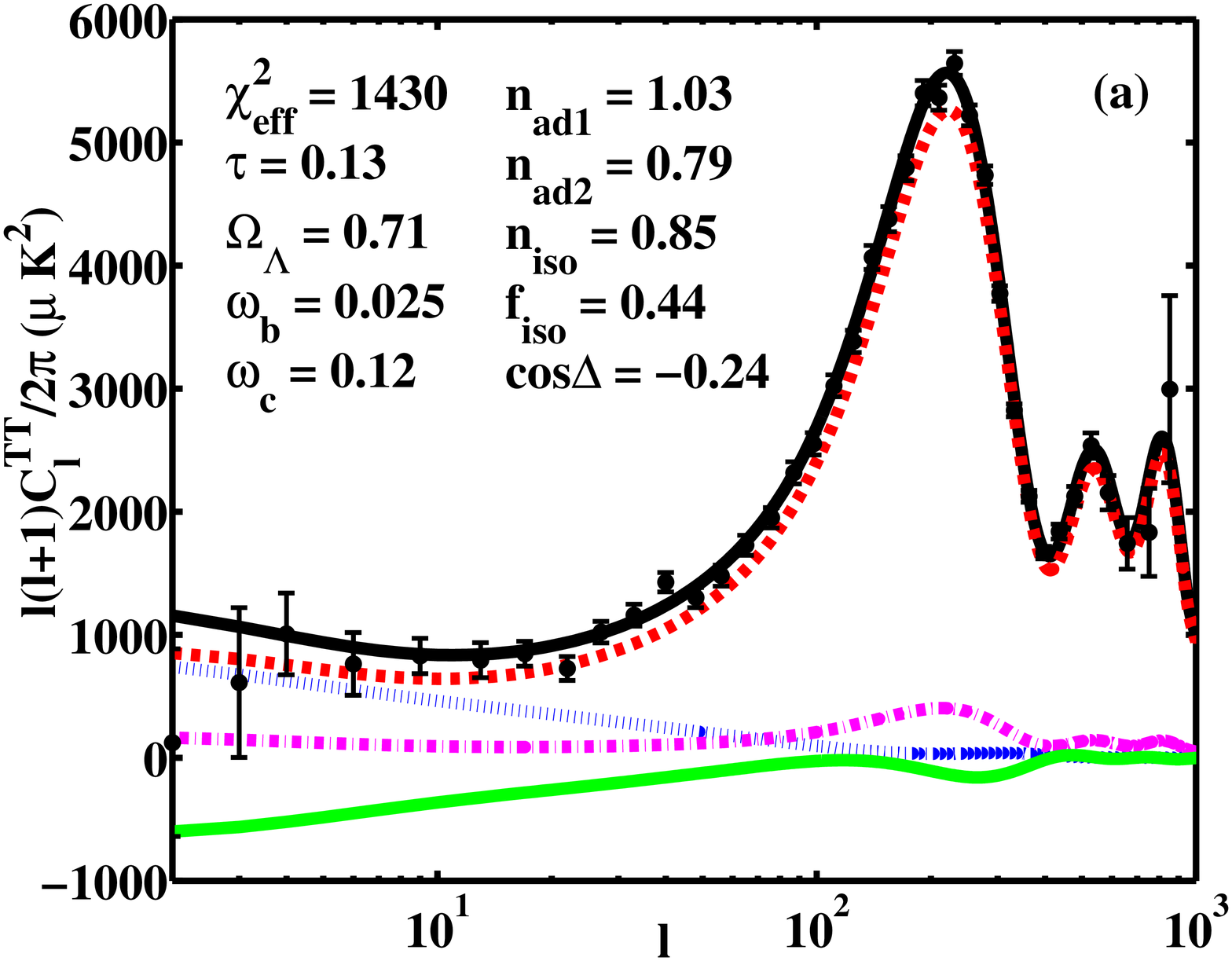}\hspace{1cm}
  \includegraphics[width=0.38\textwidth]{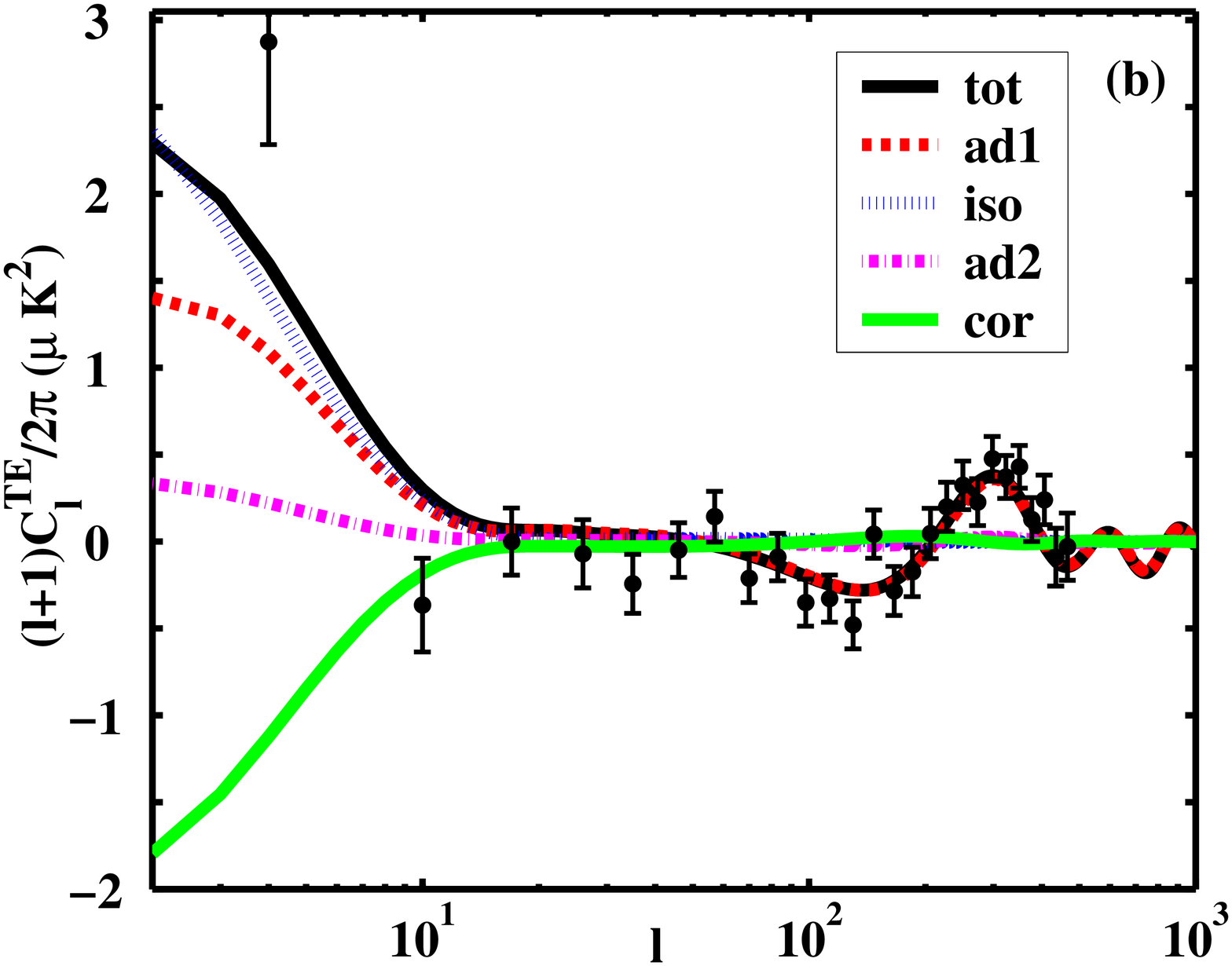}
  \caption{An example of a model that is within $2\sigma$ from our
    best-fit model.
    In the temperature power spectrum (a) the vertical axis is
    $l(l+1)C_l/2\pi$ and in the TE cross-correlation spectrum (b)
    the vertical axis is $(l+1)C_l/2\pi$!}
\end{figure*}

To demonstrate the role of the different components of the spectrum, we plot 
an angular
power spectrum of a $2\sigma$ allowed model in Fig.~2. In this particular model,
a high 64\% contribution of isocurvature to the total $C_l^{TT}$ at the quadrupole
(l=2) is allowed since the negative correlation mostly cancels the excess power.
In the TE power spectrum there is 103\% of isocurvature at the quadrupole. In
$C_l^{TE}$ the cancellation between the correlation and isocurvature is not exact
at the quadrupole, so that the isocurvature adds some power there compared to
pure adiabatic models. The \wmap measured a very high quadrupole in TE spectrum,
which was regarded as a hint for an unexpectedly large optical depth
due to reionization $\tau = 0.17\pm0.04$ \cite{WMAP1}. In positively correlated
models one would easily get a high TE quadrupole even with much smaller $\tau$.
However, positive correlation is not favored by TT spectrum where the measured
quadrupole is very low. In any case, in most models allowing for
correlation adds some power at TE quadrupole, which explains our observation
that correlated models seem to favor slightly smaller $\tau$ than pure adiabatic
models. On the other hand, strong negative correlation along with quite a large
$\tau$ might give a reasonable fit to both the low TT and the high TE quadrupole.

The isocurvature modes introduce also another degeneracy for main cosmological
parameters. Namely, allowing for general initial conditions prevents one from
determining  $\omega_b$ from the CMB \cite{Trotta:2001yw}. Nearly any value for
$\omega_b$ is allowed,
since it is determined by the relative heights of the acoustic peaks, which are also
affected by even a small isocurvature or correlation contribution. In our case
the degeneracy is even more severe than in \cite{Trotta:2001yw}, since
we allow for independently varying 
spectral indices. The big bang nucleosynthesis calculations are
valuable to determine $\omega_b$.

\begin{acknowledgments}
\paragraph{Acknowledgments.}
This work was supported by the Academy of Finland Antares Space Research
Programme grant no.~51433.  We thank H.~Kurki-Suonio, K.~Enqvist,
and H.~Ruskeep\"{a}\"{a} for comments,
M.~S.~Sloth, A.~Jokinen, A.~V\"{a}ihk\"{o}nen, J.~H\"{o}gdahl, and P.~Salmi for
discussions, and CSC -- Scientific Computing Ltd (Finland) for computational
resources.
\end{acknowledgments}

\end{document}